\documentclass[aps,prr,reprint, amsmath, amssymb,superscriptaddress]{revtex4-1}

\pdfoutput=1
\usepackage[utf8]{inputenc}
\usepackage{bm}
\usepackage[english]{babel}
\usepackage[T1]{fontenc}
\usepackage{mathrsfs}
\usepackage[retainorgcmds]{IEEEtrantools}
\usepackage{amsmath}
\usepackage{amssymb}
\usepackage{color}
\usepackage{amsfonts}
\usepackage{times,txfonts}
\usepackage{nicefrac}
\usepackage[colorlinks=true,linkcolor=blue,urlcolor=blue,citecolor=blue,pdfusetitle]{hyperref}
\usepackage{mathtools}

\usepackage{float}
\usepackage{epsfig} 
\usepackage{subfigure}
\graphicspath{{images/}}

\usepackage{physics}

\usepackage{lipsum}
\usepackage{enumerate}

\usepackage[dvipsnames]{xcolor}
\usepackage{accents}

\makeatletter
\newsavebox{\@brx}
\newcommand{\llangle}[1][]{\savebox{\@brx}{\(\m@th{#1\langle}\)}%
  \mathopen{\copy\@brx\kern-0.5\wd\@brx\usebox{\@brx}}}
\newcommand{\rrangle}[1][]{\savebox{\@brx}{\(\m@th{#1\rangle}\)}%
  \mathclose{\copy\@brx\kern-0.5\wd\@brx\usebox{\@brx}}}
\makeatother

\begin{document}

\title{First Passage Times for Continuous Quantum Measurement Currents}

\date{\today}

\author{Michael J. Kewming}
\email{kewmingm@tcd.ie}
\affiliation{School of Physics, Trinity College Dublin, College Green, Dublin 2, Ireland}
\author{Anthony Kiely}
\affiliation{School of Physics, University College Dublin, Belfield, Dublin 4, Ireland}
\affiliation{Centre for Quantum Engineering, Science, and Technology, University College Dublin, Belfield, Dublin 4, Ireland}
\author{Steve Campbell}
\affiliation{School of Physics, University College Dublin, Belfield, Dublin 4, Ireland}
\affiliation{Centre for Quantum Engineering, Science, and Technology, University College Dublin, Belfield, Dublin 4, Ireland}
\affiliation{Dahlem Center for Complex Quantum Systems, Freie Universit\"at Berlin, Arnimallee 14, 14195 Berlin, Germany}
\author{Gabriel T. Landi}
\affiliation{Department of Physics and Astronomy, University of Rochester, Rochester, New York 14627, USA}

\begin{abstract}
The First Passage Time (FPT) is the time taken for a stochastic process to reach a desired threshold. 
In this letter we address the FPT of the stochastic measurement current in the case of continuously measured quantum systems. 
We find that our approach, based on a charge-resolved master equation related to Full-Counting statistics of charge detection, enables efficient and analytical computation of the FPT. 
We develop a versatile framework applicable to quantum jump unravelling and quantum diffusion scenarios, demonstrating that the FPT can be obtained by introducing absorbing boundary conditions. Our framework is demonstrated with two relevant examples: First, we examine the tightness of recently proposed kinetic uncertainty relations (KURs) for quantum jumps, which place bounds on the signal-to-noise ratio of the FPT. Second, we investigate the usage of qubits as threshold detectors for Rabi pulses, showing how our method can optimize detection probability while minimizing false positives. 
This work offers insights into the applications of the FPT for continuous measurements including the signal-to-noise ratio bounds and false positive minimization strategies, advancing quantum information processing applications.
\end{abstract}

\maketitle

The First Passage Time (FPT), also known as the first exit, hitting, or stopping time, is a useful concept, describing the time it takes for a stochastic process to first reach a certain threshold~\cite{gillespie1991markov,gardiner1985handbook}. 
For example, if one is counting the stochastic number of particles $N(t)$ flowing into and out of a system, the FPT distribution addresses the question ``\emph{What is the probability that it takes a time $t$  until $N(t)$ first hits a specified threshold $N_{\rm th}$?}"
{This can be used in thermodynamic tasks involving Maxwell demons aimed at extracting work or cooling down a system, by stopping the dynamics whenever a certain threshold is reached~\cite{Manzano_thermodynamics_2021}.  
In the case of continuous signals, the FPT form the basis of threshold detectors~\cite{Sathyamoorthy2016,Petrovnin2023}, such as transition-edge sensors~\cite{Irwin2005}, which yield a single bit of information (``yes/no'') depending on whether a signal crosses a threshold or not.} The pervasiveness of these questions means FPTs find fertile application in a diversity of settings, in both classical and quantum systems~\cite{Roldan_decision_2015,Neri_Second_2020,Ptaszynski_first_2018,Saito_2016, Neri_statistics_2017,Garrahan_Simple_2017,Gingrich_fundamental_2017, Manzano_quantum_2019,Gianmaria_dissipation_2020, Pal_thermodynamics_2021, Vu_thermodynamics_2022, he2022quantum,Singh_universal_2019}.

{Within a quantum setting, FPTs can be formulated in terms of continuous measurements and the resulting measurement outcomes. 
Usually, these outcomes come in two flavors. For quantum jumps~\cite{carmichael_1989,Plenio1998}, they have the form of a discrete set of jump times and jump channels, while for quantum diffusion they are represented by a continuous noisy signal~\cite{wiseman2009quantum,Landi_2023_current}. 
In either case, the basic idea is the same: the continuous monitoring of the quantum system yields a classical stochastic process $X(t)$, based on which we want to create a \emph{stopping criteria} that stops the dynamics when some function of $X(t)$ crosses a certain threshold value. A particular case of this problem is that of}
waiting time distributions (WTDs), which have been explored in classical stochastic processes~\cite{stratonovich_book}, quantum optics \cite{vyas1988,carmichael_1989}, electronic transport \cite{brandes_2008,thomas2013,haack2014},  thermodynamics~\cite{skinner2021,Manzano_thermodynamics_2021,Garrahan_Simple_2017,Vu_thermodynamics_2022}, and condensed matter~\cite{schulz2022}. 
The WTD describes the statistics of the time between two events {while FPTs describe the time until an arbitrary number of events accumulate to reach a certain threshold. 
WTDs are therefore a particular case of FPTs.}
There has been a significant body of work in WTDs for continuously measured quantum systems~\cite{carmichael_1989,Landi_2023_current,brandes_2008,brandes2016,kosov2016,ptaszynski2017,kleinherbers2021,vyas1988,albert2011,thomas2013,Stefanov2022}, 
{which is by now well understood and  relatively easy to compute.
There has also been} some earlier work focused on computing the FPT in homodyne and heterodyne measurements for two level emitters \cite{Anders_stochastics_2014} {and solid state qubits~\cite{Korotkov2006}}.
However, a general description for FPT is still lacking.
As a consequence, the only way of computing them is through expensive statistical (Monte Carlo) sampling over various quantum trajectories. 
{This is not only extremely costly from a computational point of view, but also lacks any analytical insights.} A more systematic methodology {that is able to deterministically compute FPTs would therefore be quite valuable.}

In this work we address this deficiency and derive a method for deterministically computing the FPT distribution {for stopping criteria based on the net accumulated current $N(t)$ through a continuously measured system}.
We first show how the unconditional evolution can be decomposed in terms of a \emph{charge-resolved dynamics}. This is a concept already explored in specific contexts, such as quantum optics~\cite{Zoller_quantum_1987,Plenio1998} and mesoscopic transport~\cite{Li_Spontaneous_2005, li2016number}. 
Here, we show more generally that it can be formulated as the Fourier transform of the generalized (tilted) master equation used in Full-Counting Statistics (FCS)~\cite{Landi_2023_current,Landi2022,schaller2014,esposito2009,levitov1993,levitov1996,nazarov2003,flindt2010}, which we use to establish a charge resolved equation for both the quantum jump and the quantum diffusion unravellings. 
Armed with this dynamics, we then show how the FPT problem can be implemented by  imposing absorbing boundary conditions.
We apply our results to characterize the tightness of recently developed Kinetic Uncertainty Relations (KURs) for FPTs~\cite{Garrahan_Simple_2017,Vu_thermodynamics_2022}. 
{We also study the diffusive measurement of a qubit's population and show how this can be used as a threshold detector for the application of Rabi pulses.}

{\it First passage times.---}Consider a generic stochastic process $X(t)$ (either continuous- or discrete-space) governed by a probability distribution $P(x,t)$, and starting at $X(0)=x_0$. 
We make no assumption about the kind of dynamical equation that $P(x,t)$ obeys. 
All we assume is that there exists a rule taking $P(x,t) \to P(x,t+dt)$. 
Given a certain region $\mathcal{R} = [a,b]$ (assuming $x_0 \in \mathcal{R})$, the FPT is the random time $\tau$ at which $X(t)$ \emph{first} leaves $\mathcal{R}$.
The most effective way of computing this is by imposing absorbing boundary conditions. 
That is, at each time-step of the evolution we impose that $P(x,t) = 0$ for all $x\notin \mathcal{R}$. 
This causes $P(x,t)$ to evolve differently, giving rise to a new distribution $P_\mathcal{R}(x,t)$, which is no longer normalized. 
The normalization constant is the survival probability $G_\mathcal{R}(t) = \int_a^b dx P_\mathcal{R}(x,t)$ {that $X(t)$ is still in $\mathcal{R}$ at time $t$}.
The probability density that the threshold is first crossed at time $t$ is the FPT distribution~\cite{gillespie1991markov, gardiner1985handbook}:
\begin{equation}
\label{eq:absorbing_classic}
    f_{\mathcal{R}}(t)=
    - \frac{d G_{\mathcal{R}}(t)}{d t}\,.
\end{equation}
If $G_\mathcal{R}(\infty) = 0$ the boundary is always eventually reached, and consequently $\int_0^\infty f_\mathcal{R}(t) = 1$. 
But this need not always be the case. 

{\it FPT from continuous quantum measurements.---}We replace the random variable $X(t)$ with an integrated current, $N(t)$, corresponding to the output of some continuous measurement detector. 
For example, $N(t)$ could be the total number of detected photons from a leaky optical cavity~\cite{vyas1988,carmichael_1989}, the net-particle current from a thermal machine~\cite{Karimi2020,Manzano_thermodynamics_2021}, the continuous diffusive readout of a resonator coupled to a superconducting circuit~\cite{Murch2013,Campagne_Ibarcq2016,Didier_2015,Blais_2021,he2022quantum,Naghiloo2020}, or a continuous charge measurement from a quantum point contact~\cite{Hofmann2016}. 
We assume that the system evolves unconditionally according to a Lindblad master equation (with $ \hbar = 1$)
\begin{equation}
\label{eq:liouv}
    \frac{d\rho}{dt} =\mathcal{L}\rho= - i [H, \rho] + \sum_{k} L_{k}\rho L_{k}^{\dagger} - \frac{1}{2}\{L_{k}^{\dagger}L_{k}, \rho \}\,,
\end{equation}
where $H$ is the Hamiltonian and $L_k$ represent different jump channels.
We separately treat the quantum jump and quantum diffusion unravellings. 
In each case, we also detail how the physical currents are constructed from the output data.

\emph{Jump unraveling -} In this case the measurement outcomes at each interval $dt$ are random variables $dN_k(t) = 0$ or $1$, taking the value $1$ whenever there is a jump in channel $k$, which occurs with probability $dt \tr\big\{ L_k^\dagger L_k \rho_c(t)\big\}$.
The conditional density matrix $\rho_c(t)$,  given the  measurement outcomes,  evolves according to the It\^o stochastic master equation~\cite{wiseman2009quantum, Landi_2023_current} 
\begin{equation}\label{stochastic_jump}
    d \rho_{c}(t) =  \Big\lbrace\sum_{k}dN_{k}(t) \mathcal{G}[L_{k}] - dt \mathcal{H}\left[ iH_{\rm eff} \right]\Big\rbrace \rho_{c}(t)\,,
\end{equation}
where $\mathcal{G}[A]\rho = A\rho A^{\dagger}/\langle A^{\dagger}A\rangle_c - \rho$ and  $\mathcal{H}[A]\rho = A\rho + \rho A^{\dagger} - \langle A + A^{\dagger}\rangle_c \rho$, with $\langle \bullet  \rangle_c = \tr\{ \bullet  \rho_c(t)\}$ and the effective Hamiltonian $H_{\rm eff} = H - \frac{i}{2}\sum_k L_k^\dagger L_k$. 
The stochastic charge up to time $t$ can be defined  generally as 
$N(t) = \sum_k \nu_k \int dN_k(t)$,
where $\nu_k$ are problem-specific coefficients describing the physical current in question. 
For example, in a system with one injection channel $L_+$ and one   extraction channel $L_-$, the excitation current would have $\nu_+ = +1$ and $\nu_- = -1$, leading to a net charge $N(t) = \int \big[ dN_+(t) - dN_-(t)\big]$.

The main object in FCS is the  distribution $P(N,t)$ giving the probability that the stochastic charge $N(t)$ has a value $N$ at time $t$. 
Here we utilize the concept of a \emph{charge-resolved density matrix} $\rho_N(t)$, defined such that $P(N,t) {=} \tr\{\rho_N(t)\}$ and $\sum_N \rho_N(t) {=} \rho(t)$ (the unconditional state).
The charge-resolved density matrix was first introduced for monitoring quantum jumps in quantum optics \cite{Zoller_quantum_1987} and later extended to quantum transport in mesoscopics \cite{li2016number, Li_Spontaneous_2005}. 
In terms of the conditional dynamics it reads $\rho_N(t) {=} E\big[\rho_c(t) \delta_{N(t),N} \big]$, where $N(t)$ is the stochastic charge and $E[\bullet]$ refers to the ensemble average over all trajectories~\footnote{As a consistency check, note how
$\rho(t) = \sum_{N} ~\rho_N(t) = \sum_N~E\Big[\rho_c(t) \delta_{N(t), N} \Big] = E\big[ \rho_c(t)]$, as expected
}. {Hence $\rho_N(t)$ can be interpreted on the ensemble averaged dynamics, conditioned on the assumption that at time $t$ the total accumulated charge is $N$.}
Using the tilted Liouvillian from FCS~\cite{Landi2022,Landi_2023_current,schaller2014,esposito2009}, we show in \cite{supp_material} that $\rho_N(t)$ evolves according to the charge-resolved equation
\begin{equation}
    \label{eq:nresolvedkraus}
    \frac{\partial \rho_N}{\partial t} = \mathcal{L}_0 \rho_N + \sum_k L_k \rho_{N-\nu_k} L_k^\dagger,
\end{equation}
where $\mathcal{L}_0 \rho = - i (H_{\rm eff} \rho - \rho H_{\rm eff}^\dagger)$ is the no-jump evolution. 
The initial condition is 
$\rho_N(0) = \delta_{N,0} \rho(0)$.
Eq.~\eqref{eq:nresolvedkraus} is system of coupled master equations for each density matrix $\rho_N$.
The first term describes how each $\rho_N$ changes due to a no-jump trajectory, while the other terms describe how $\rho_N$ connect with $\rho_{N-\nu_k}$ through the jump channel $L_k$. 

Eq.~\eqref{eq:nresolvedkraus} can now be adapted to yield the FPT statistics for $N(t)$ to 
leave a certain pre-defined boundary $\mathcal{R} = [a,b]$ (with $a<0$ and $b>0$). 
We do this by imposing absorbing boundary conditions, $\rho_{N<a}(t)=\rho_{N>b}(t) = 0$.
This causes the system to follow a modified evolution $\rho_N^\mathcal{R}(t)$ from which we obtain $P_\mathcal{R}(N,t) = \tr \big\{ \rho_N^\mathcal{R}(t)\big\}$~\footnote{Notice that if $a\to -\infty$ and $b\to \infty$, we recover the standard FCS distribution $P(N,t)$}. 
The survival probability is then $G_\mathcal{R}(t) = \sum_{N=a}^b P_\mathcal{R}(N,t) = \sum_{N=a}^b \tr\big\{\rho_N^\mathcal{R}(t)\big\}$.
Differentiating with respect to time using 
Eqs.~\eqref{eq:absorbing_classic} and~\eqref{eq:nresolvedkraus}, we obtain the FPT 
\begin{equation}\label{FPT1}
    f_\mathcal{R}(t) = \sum_{N=a}^b \sum_k \tr\big\{ L_k^\dagger L_k (\rho_N^\mathcal{R} - \rho_{N-\nu_k}^\mathcal{R})\big\}.
\end{equation}
To arrive at this result we also used the fact that $\mathcal{L}_0 \rho = \mathcal{L}\rho - \sum_k L_k\rho L_k^\dagger$, as well as the fact that $\tr\big\{ \mathcal{L}(\bullet)\big\} = 0$. 
Eq.~\eqref{FPT1}, together with~\eqref{eq:nresolvedkraus}, form our first main result.
They connect the FPT directly to the solution of the charge-resolved master equation {and the probabilities of charge flowing out of $\mathcal{R}$.}
In the case of two jump operators $L_\pm$ with $\nu_k = \pm 1$ Eq.~\eqref{FPT1} simplifies to
$f_\mathcal{R}(t) = \tr\big\{ L_+^\dagger L_+ \rho_b^\mathcal{R}(t)\big\} +
    \tr\big\{ L_-^\dagger L_- \rho_a^\mathcal{R}(t)\big\}$.
This shows that all that matters are the states at the boundaries of the region $\mathcal{R}$. The two terms can be interpreted as \emph{conditional escape rates} for $N(t)$ to leave $\mathcal{R}$, given it has not yet done so up to time $t$.
The above results provide a deterministic and efficient method to obtain $f_\mathcal{R}(t)$.
Not only does it avoid sampling over quantum trajectories, but Eq.~\eqref{eq:nresolvedkraus} is also just a linear system of equations for the variables $\rho_N^{\mathcal{R}}$.
In fact, in terms of a vector $\vec{\rho}^{\,\mathcal{R}} = (\rho_a^\mathcal{R},\rho_{a+1}^\mathcal{R},\ldots,\rho_b^\mathcal{R})$, Eq.~\eqref{eq:nresolvedkraus} reduces simply to $\partial_t \vec{\rho}^{\,\mathcal{R}} = \mathcal{V} \vec{\rho}^{\,\mathcal{R}}$, for a superoperator $\mathcal{V}$ \cite{supp_material}.

\emph{Diffusion unraveling - }
The diffusive unraveling of Eq.~(\ref{eq:liouv}) is written as the It\^o stochastic differential equation~\cite{Landi_2023_current,wiseman2009quantum}
\begin{equation}\label{stochastic_diffusion}
    d\rho_{c}(t) = dt \mathcal{L} \rho_c(t) + \sum_{k}\mathcal{H}[L_{k}e^{-i\phi_{k}}]\rho_{c}dW_{k}(t),
\end{equation}
with independent Wiener increments $dW_k(t)$.
The current and charge in this case are given by 
\begin{equation}\label{diffusion_current}
    I(t) =\sum_k \nu_k \Bigg(\langle x_k \rangle_c + \frac{dW_k}{dt}\Bigg), 
    \qquad 
    N(t) = \int\limits_0^t dt' I(t'),
\end{equation}
where $x_k = L_k e^{-i \phi_k} + L_k^\dagger e^{i \phi_k}$ (with $\phi_k$ being arbitrary angles).
The charge $N(t)$ is now a continuous stochastic process. 
Notwithstanding, we can similarly define a charge-resolved density matrix $\rho_N$.
The equation for $\rho_N$ is derived in \cite{supp_material} using the tilted Liouvillian of quantum diffusion recently derived in~\cite{Landi_2023_current}. 
The result is 
\begin{equation}
    \label{eq:nresolveddiff}
    \frac{\partial \rho_{N}(t)}{\partial t} = \mathcal{L}\rho_{N}(t) - \sum_{k}\nu_{k}\mathcal{K}[L_{k}e^{-i\phi_{k}}]\frac{\partial \rho_{N}(t)}{\partial N}+ \frac{K_{\rm diff}}{2}\frac{\partial^{2}\rho_{N}(t)}{\partial N^{2}}\,,
\end{equation}
where $\mathcal{K}[A]\rho = A\rho + \rho A^{\dagger}$ and $K_{\rm diff} = \sum_k \nu_k^2$ is a constant.
Once again, we obtain the FPT by imposing absorbing boundary conditions $\rho_{N<a}(t) = \rho_{N>b}(t) \equiv 0$.
Eq.~\eqref{eq:nresolveddiff}, which is a type of quantum Fokker-Planck equation~\cite{Annby_Andersson2022}, is our second main result. 

{\it Kinetic uncertainty relation (KUR).---}From $f_\mathcal{R}(t)$ we can compute the average FPT $E[\tau]$ and its variance ${\rm Var}[\tau]$.
{Of particular interest is} the signal-to-noise ratio (SNR)
${\rm SNR}_{\tau} = \frac{{\rm E} [\tau]^{2}}{{\rm Var}[\tau]}$.
{For instance, in the context of autonomous clocks, this quantity is related to the timekeeping precision~\cite{Erker2017} and was recently studied experimentally~\cite{he2022quantum} in superconducting circuits, showcasing the non-trival role of quantum coherence.} 
In Ref.~\cite{Garrahan_Simple_2017} it was proven that for classical (or incoherent) systems the SNR is bounded by 
${\rm SNR}_{\tau} \leqslant E[\tau] K$, where $K = \sum_k \tr\big\{ L_k^\dagger L_k \rho_{\rm ss}\big\}$ is the dynamical activity (number of jumps per unit time) and $\rho_{\rm ss}$ is the steady-state of~\eqref{eq:liouv}.
This bound, however, can be violated for coherent dynamics.
Motivated by that, Ref.~\cite{Vu_thermodynamics_2022} derived the bound 
${\rm SNR}_{\tau} \leqslant E[\tau] (K+\mathcal{Q})$, where $\mathcal{Q}$ is a quantum correction \cite{supp_material}. 

\begin{figure}
    \centering
    \includegraphics[width=\columnwidth]{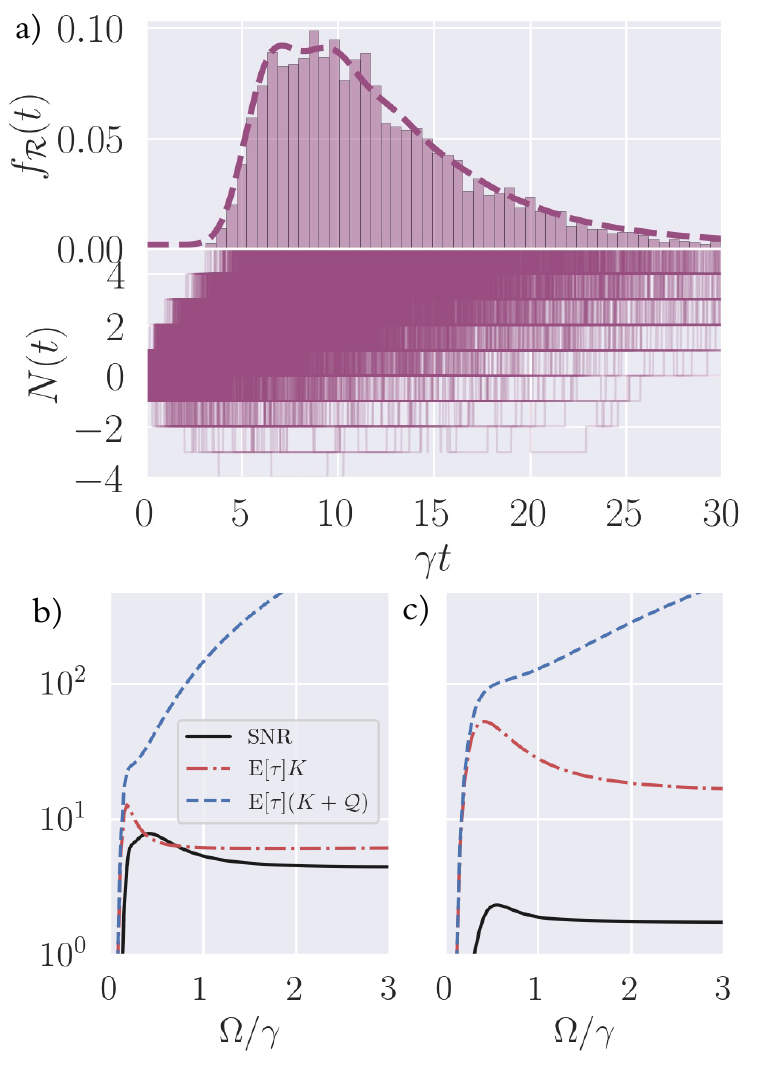}
    \caption{KURs for a driven qubit. a) FPT distribution for $N(t)$ crossing $\mathcal{R} = (-\infty,5]$, with $\Omega = \gamma = 1$, $\bar{n} = 0.2$ and initial state $\rho_{ss}$. The histogram corresponds to the quantum trajectory simulations, depicted below as a function of time.
    b,c) SNR (black solid),  classical (KUR)~\cite{Garrahan_Simple_2017} (red `-- $\cdot$') and quantum KUR~\cite{Vu_thermodynamics_2022} (blue `-') as a function of $\Omega/\gamma$, for $\bar{n}=0.1$ and $1$ respectively. In b) we can see a violation of the classical bound.
    }
    \label{fig:jumps}
\end{figure}

A relevant open question concerns the tightness of these bounds. 
This can be difficult to address because computing the SNR requires sampling over many quantum trajectories. 
Eq.~\eqref{eq:nresolvedkraus}, however, makes this task straightforward.
Here we illustrate this idea by considering a resonantly driven qubit with rotating frame Hamiltonian  
$H = \Omega \sigma_x$, 
where $\sigma_{\alpha}$ are Pauli matrices
and $\Omega$ is the strength of the Rabi drive.
We further assume that this  is immersed in a thermal environment  described by the Lindblad master equation~\eqref{eq:liouv} {with jump operators
$L_- {=} \sqrt{\gamma(\bar{n}+1)} \sigma_-$ and $L_+ {=} \sqrt{\gamma\bar{n}} \sigma_+$, where 
$\gamma$ is the decay rate and $\bar{n}$
is the Bose-Einstein occupancy.
}
We focus on the excitation current, by defining $\nu_\mp = \pm 1$. 
For the purpose of illustration, we also set $\mathcal{R} = (-\infty, 4]$.
An example of $f_{\mathcal{R}}(t)$, computed using Eq.~\eqref{eq:nresolvedkraus}, is shown in Fig.~\ref{fig:jumps}(a) in dashed lines. 
We also compare our results with quantum trajectories, where the FPT is obtained by histogramming the times at which the charge reaches the threshold $N(t) = 5$ in each trajectory.

We next use this model to study the KURs in Refs.~\cite{Garrahan_Simple_2017,Vu_thermodynamics_2022}.
Formulas for $K$ and $\mathcal{Q}$ are given in \cite{supp_material}.
Results comparing the SNR with the two bounds are shown in Fig.~\ref{fig:jumps}(b,c) for $\bar{n} = 0.1$ and $1$.
We see that the classical bound~\cite{Garrahan_Simple_2017} is somewhat tight, and tends to follow the overall behavior of the SNR. 
However, it can be violated, as in Fig.~\ref{fig:jumps}(b). 
Such quantum violations have been the subject of extensive research~\cite{Agarwalla2018,Ptaszy_ski2018,Liu2019,Saryal2019,Cangemi2020,Kalaee2021,Prech2023} as they are connected to dynamical aspects of  coherence.
Conversely, the quantum bound of~\cite{Vu_thermodynamics_2022} is never violated, as it must. 
However, it is also rather loose and diverges as $\mathcal{Q} \propto \Omega^2$.
This result is relevant for the following reason. 
The classical bound depends only on the dynamical activity $K$ of the observed quantum jumps associated to the operators $L_k$. 
But in quantum coherent problems there is also activity associated to the unitary dynamics (the Rabi oscillations in our case), although this is hidden to the observer.
This additional activity is precisely what $\mathcal{Q}$ captures. 

{{\it Single-qubit threshold detector.---}As our second application, we use our framework to model a single qubit functioning as a threshold detector for Rabi pulses. This is motivated by the recent experiment in Ref.~\cite{Petrovnin2023}.
Suppose one wishes to know whether a Rabi pulse $\Omega(t)$ (of unknown shape and duration) was applied to a qubit during some time window $\tau$.
The goal is to come up with a ``yes/no'' protocol, based on a continuous measurement record of the qubit, that yields ``yes'' (click) if the pulse was applied and ``no'' (no-click) if it was not. 
To do that, we continuously monitor the qubit's population, within the diffusive unravelling, resulting in a stochastic net charge $N(\tau)$ [Eq.~\eqref{diffusion_current}]. 
We then choose the interval $\mathcal{R}=[a,b]$ and associate $N(\tau) \notin \mathcal{R}$ with a click (the  pulse was applied), and $N(\tau) \in \mathcal{R}$ with no click (the pulse was not applied).
In this way, the threshold detector is cast as a first passage time problem. 
The goal is to choose $\tau$ and $\mathcal{R}$ in order to maximize the successful detection probability $p_{\rm succ}$ and, at the same time, minimize the probability of false positives $p_{\rm false}$  (when the detector clicks ``yes'' even though no pulse was applied). 

We model this using Eq.~\eqref{stochastic_diffusion} with $H(t) = \Omega(t) \sigma_x$  and a single jump operator $L = \sqrt{\gamma}\sigma_z$. We choose $\nu = \sqrt{\gamma}$ to make $N(\tau)$ in Eq.~\eqref{diffusion_current} dimensionless. 
The shape and structure of $\Omega(t)$ depend on the pulse in question.
We assume the system starts in $|\!\!\downarrow\rangle$, so if $\Omega(t) \equiv 0$, it will remain there throughout. 
Any Rabi pulse will therefore tend to partially excite the qubit, which in turn will change the stochastic properties of the signal $N(\tau)$.
For a given $\Omega(t)$, the detection probability is obtained by solving Eq.~\eqref{eq:nresolveddiff}, with initial state $|\psi_0\rangle = |\!\!\downarrow\rangle$, and following the same steps delineated before to compute the survival probability $G_\mathcal{R}(\tau|\Omega(t), |\!\!\downarrow\rangle)$. 
The successful detection probability is then  $p_{\rm succ} = 1- G_\mathcal{R}(\tau|\Omega(t), |\!\!\downarrow\rangle)$. 
Conversely, the false positive probability is 
$p_{\rm false} = 1- G_\mathcal{R}(\tau|\Omega(t)\equiv 0, |\!\!\downarrow\rangle)$.}

\begin{figure}
    \centering
    \includegraphics[width=0.4\textwidth]{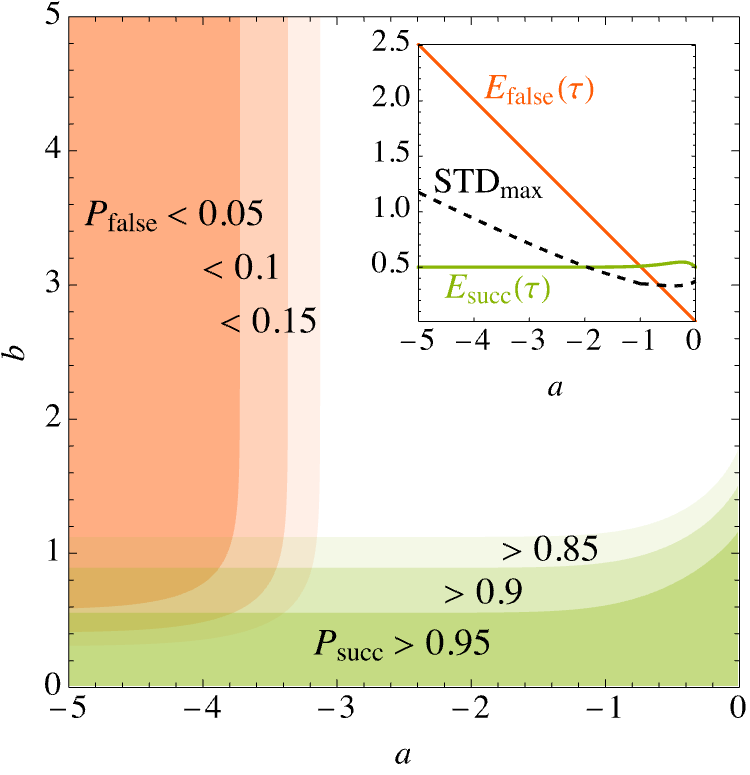}
    \caption{{Operation of a continuously monitored qubit as a threshold detector. 
    The plot shows regions in the $(a,b)$ plane with bounds on the false positive detection probability $p_{\rm false}$ and the successful detection probability $p_{\rm succ}$, with fixed $\gamma \tau = 1$ and $q = 1$ (analytical expressions are shown in \cite{supp_material}.
    The inset shows the average FPT, in units of $\gamma$, for the false positive and the successful detection cases respectively, with $b=1$ and $q=1$. It also shows the maximum possible standard deviation of the FPT, obtained by maximizing over all $q$. }}
    \label{fig:threshold}
\end{figure}

{The probability $p_{\rm succ}$ depends on the specifics of $\Omega(t)$. 
For concreteness and simplicity, we will focus here on a delta-like pulse  $\Omega(t) = \Omega_0 \delta(t)$.
The complete analytical solution can be found in \cite{supp_material}. 
The resulting success probability reduces to 
$p_{\rm succ} = p_- + q (p_+ - p_-)$ where $q = \langle \sigma_+\sigma_-\rangle_0 = \sin^2(\Omega_0)$ is the initial state occupation and $p_\pm$, which depend only on $a,b,\gamma\tau$, are the detection probabilities for initial states $|\!\!\uparrow\rangle$ and $|\!\!\downarrow\rangle$, respectively. 
The false positive probability is 
$p_{\rm false} = p_-$.
The goal is to minimize $p_{\rm false}$ and maximize $p_{\rm succ}$. 
Fig.~\ref{fig:threshold} shows regions in the $(a,b)$ plane representing different bounds on $p_{\rm false/succ}$, for fixed $\gamma \tau = 1$ and $q=1$. 
From this image one can infer that optimal operation occurs for small $b$ and large $|a|$; e.g. $b\sim 1$ and $a\sim -5$. 
Similar conclusions can be drawn by looking at the mean and variance of the FPT. 
The mean for the two processes are plotted on the inset of Fig.~\ref{fig:threshold} as a function of $a$, with $b = 1$ and $q=1$.
For large $|a|$, $E_{\rm false}(\tau) \gg \gamma\tau = 1$, so false positives are unlikely to occur for this value of $\gamma\tau$. 
The inset of Fig.~\ref{fig:threshold} also shows how the maximum standard deviation of the FPT, maximized over all $q$, does not grow significantly with $a$.  
So not only are false positives unlikely on average, but their fluctuations is also small. 
These results, combined, corroborate this parameter regime as useful for the operation as a detector. 
Of course, this analysis pertains only to a toy model and, in reality, several other factors would have to be taken into consideration. 
Notwithstanding, they serve to illustrate how, through  the analytical insights from our framework, one can systematically search for optimal operating regimes.
}

{\it Discussion and conclusions.---}Our methodology is compatible with any type of master equation in the form~\eqref{eq:liouv}, including time-dependent Hamiltonians.
It therefore encompass a broad range of physical problems, from quantum optics to condensed matter. 
In addition, because Eqs.~\eqref{eq:nresolvedkraus} and~\eqref{eq:nresolveddiff} are resolved in $N$, it is straightforward to extend our method to incorporate $N$-dependent feedback.
That is, to study models where $H$ or $\{L_k\}$ are modified depending on the current value of $N(t)$ in a quantum trajectory~\cite{Annby_Andersson2022}. 
Despite not being the  focus of this letter, we emphasize this connection because feedback and FPTs are actually conceptually very similar: both require monitoring of a stochastic quantity and performing (or not) actions depending on its value. In the case of the FPT, the action is to continue or cease the dynamics. In the case of feedback, it is to modify the Liouvillian. 
Feedback and FPT also share the same practical difficulty of requiring computationally expensive quantum trajectories. 
Deterministic strategies, such as the one put forth in this letter, are therefore crucial. 
A famous example of a successful deterministic theory is that of current feedback put forth in Refs.~\cite{Wiseman1994,Wiseman1994b}, which had a significant impact, despite working only for a restricted class of models. We believe a similar point can be made for our results. 



A particularly interesting application of our results is to  so-called gambling problems, such as that studied in~\cite{Manzano_thermodynamics_2021}. 
This involves an agent which uses information about the system's state to devise stopping strategies aimed at maximizing a certain goal, which can be relevant in the context of thermodynamics.
For instance, depending on the model $N(t)$ can be related to the heat exchanged with the bath, or the work performed by an external drive.
An agent with access to either of these quantities could then devise a strategy such as ``stop the process whenever a certain amount of work has been extracted.''
This has interesting thermodynamic implications, as it puts in the foreground the role of information in thermodynamic processes. 
It may also have practical consequences. 
For example, one can use these ideas to devise optimal cooling protocols, or for quantum state engineering.

\emph{Acknowledgements--} The authors thank Mark Mitchision for the helpful discussions. MJK acknowledges the financial support from a Marie Sk\l odowska-Curie Fellowship (Grant No. 101065974).
SC acknowledges support from the Science Foundation Ireland Starting Investigator Research Grant “SpeedDemon” No. 18/SIRG/5508, the John Templeton Foundation Grant ID 62422, and the Alexander von Humboldt Foundation.

\bibliography{references}
\bibliographystyle{apsrev4-1}

\end{document}